\documentclass[conference]{IEEEtran}
\IEEEoverridecommandlockouts
\usepackage{cite}
\usepackage{amsmath,amssymb,amsfonts}
\usepackage{algorithmic}
\usepackage{graphicx}
\usepackage{textcomp}
\usepackage{xcolor}

\def\BibTeX{{\rm B\kern-.05em{\sc i\kern-.025em b}\kern-.08em
    T\kern-.1667em\lower.7ex\hbox{E}\kern-.125emX}}
\begin{document}

\title{Hybrid Analytical-Machine Learning Framework for Ripple Factor Estimation in Cockcroft-Walton Voltage Multipliers with Residual Correction for Non-Ideal Effects \\
{\footnotesize }

}

\author{\IEEEauthorblockN{Md. Tanvirul Islam}
\IEEEauthorblockA{\textit{Department of Electrical and Electronic Engineering} \\
\textit{Bangladesh University of Engineering and Technology}\\
Dhaka, Bangladesh \\
Email: 2006095@eee.buet.ac.bd}

}

\maketitle

\begin{abstract}
Cockcroft-Walton (CW) voltage multipliers suffer from output ripple that classical analytical models underestimate due to neglected non-idealities like diode drops and capacitor ESR, particularly in high-stage, low-frequency and heavy-load regimes. This paper proposes a hybrid framework that generates a comprehensive 324-case MATLAB/Simulink dataset varying stages (2-8), input voltage (5-25 kV), capacitance (1-10 $\mu$F), frequency (50-500 Hz) and load (6-60 M$\Omega$), then trains a Random Forest model to predict residuals between simulated and theoretical peak-to-peak ripple. The approach achieves 70.6$\%$ RMSE reduction (131 V vs. 448 V) globally and 66.7$\%$ in critical regimes, with near-zero bias, enabling physically interpretable design optimization while outperforming pure ML in extrapolation reliability.

\end{abstract}


\section{Introduction}
Cockcroft–Walton (CW) voltage multipliers are the most popular mechanism to generate high-voltage DC outputs from low or moderate-voltage AC sources for a long time. These circuits are highly used in applications such as particle accelerators, X-ray generators, electrostatic precipitators and high-voltage test equipment. The reason behind their popularity arises from their simple structure, modular scalability and the ability to achieve very high output voltages without bulky transformers [1-3]. However, a fundamental limitation of CW multipliers is the presence of output voltage ripple and voltage drop under load, which degrades regulation and restricts their suitability for precision high-voltage applications.

 The ripple voltage in CW multipliers originates from the periodic charging and discharging of stage capacitors through rectifying diodes during each AC cycle. Classical analytical models describe ripple and voltage drop as functions of load current, operating frequency, capacitance and the number of stages [1,4]. These expressions are typically derived under idealized assumptions such as identical stage capacitances, ideal diodes and steady-state operation. As a result, analytical ripple formulas are generally conservative and represent worst-case estimates rather than realistic operating behavior, particularly under light or moderate loading conditions [5].

 In addition to idealized circuit assumptions, practical Cockcroft–Walton multipliers are affected by non-ideal component characteristics, particularly the equivalent series resistance (ESR) of stage capacitors. ESR introduces additional voltage ripple and loss due to resistive voltage drops associated with ripple current flow, which becomes increasingly significant at higher load currents and switching frequencies. Classical ripple models often neglect ESR effects, leading to further deviation between analytical predictions and realistic operating behavior, especially in multi-stage configurations where cumulative losses become non-negligible. Accurate ripple estimation therefore requires consideration of ESR alongside conventional parameters such as load resistance, capacitance, frequency, and stage number. 

 In practical systems, ripple characteristics are strongly influenced by multiple interacting parameters, including load resistance, stage count, capacitance values and excitation frequency. The combined nonlinear dependence of ripple on these parameters makes accurate analytical prediction difficult, especially when transient settling effects and post-steady-state behavior are considered [6]. Time-domain circuit simulation provides a more realistic representation of CW multiplier behavior. However, systematic exploration of large multi-parameter design spaces using simulation alone can be computationally expensive and analytically opaque. 

 Recent advances in machine learning (ML) have enabled data-driven modeling of complex nonlinear systems using simulation- or measurement-based datasets. ML techniques have been successfully applied in power electronics and high-voltage engineering for parameter estimation, performance prediction, and surrogate modeling, offering improved accuracy over traditional closed-form expressions while retaining computational efficiency [7–9]. Despite this progress, the application of ML to ripple factor estimation in CW voltage multipliers remains relatively unexplored.

This paper introduces a hybrid analytical-machine learning framework for multi-parameter ripple factor estimation in Cockcroft-Walton voltage multipliers that accounts for non-ideal effects. Classical ripple theory establishes interpretable baselines, created by a comprehensive parametric simulation dataset under varied loading, frequency, capacitance, stages and input voltage conditions. Steady-state ripple metrics and higher-order waveform features (skewness, kurtosis, crest factor) train a Random Forest residual predictor, delivering corrected estimates with 70$\%$ error reduction across regimes—especially high-stage (N$\geq$6), low-frequency (f$\leq$100 Hz), heavy-load (R$\geq$12 M$\Omega$), and critical combinations, while revealing design insights like frequency's superiority over capacitance for mitigation.

\section{Theory}
\subsection{Operating Principle of the Cockroft-Walton Multiplier}
The Cockcroft–Walton (CW) voltage multiplier is a diode and capacitor-based HVDC generation topology. This is widely used in applications requiring high output voltage at relatively low current levels. The circuit operates by charging capacitors during alternate half-cycles of an AC input voltage. By stacking these charged capacitors in series, voltage multiplication is achieved. Due to its modular structure, absence of magnetic components and ease of high-voltage insulation, the CW multiplier remains a preferred solution in many high-voltage engineering applications [1]. 

For an N-stage half-wave CW multiplier driven by an input voltage with peak amplitude $V_{in}$, the ideal no-load output voltage is expressed as, 
\begin{equation}
    V_{out, ideal}= 2NV_{in}
    \label{eq: Vout}
\end{equation}
But, in practical operation, the output voltage deviates from this ideal value due to load current, finite capacitance and non-ideal charge transfer mechanisms. This phenomenon leads to voltage drop and ripple at the output [2].

\subsection{Origin of Output Voltage Ripple}
Output voltage ripple in a CW multiplier arises from the periodic charging and discharging of the stage capacitors under load conditions. When a load current $I_L$ is drawn, charge is removed from the upper-stage capacitors during each cycle. This extracted charge must be replenished during subsequent input cycles, resulting in a time-varying output voltage.

Unlike single-stage rectifier circuits, ripple in a CW multiplier accumulates along the cascade. Higher-stage capacitors experience larger voltage fluctuations due to cumulative charge transfer through preceding stages. As a result, ripple magnitude increases with the number of stages and becomes strongly dependent on operating frequency and capacitance values [2, 3].

\subsection{Classical Analytical Ripple Model}
Analytical ripple models for CW multipliers are typically derived under the following assumptions: ideal diodes, identical stage capacitances, negligible parasitic elements, steady-state periodic operation and constant load current. Under these conditions, the peak-to-peak output ripple voltage for an N-stage CW multiplier operating at frequency f with stage capacitance C is approximated as [3],
\begin{equation}
    V_{r, pp}^{theory}=\frac{I_L}{fC}\cdot \frac{N(N+1)}{2}
    \label{eq02}  
\end{equation}
This expression indicates that ripple voltage is directly proportional to the load current and increases quadratically with the number of stages, while being inversely proportional to the product of frequency and capacitance. As a result, ripple becomes a dominant performance-limiting factor in high-stage multipliers operating at low frequency or under heavy loading conditions.

\subsection{Ripple Factor Definition}
Ripple performance is commonly quantified using the ripple factor, defined as the ratio of the RMS value of the ripple component to the average DC output voltage.
\begin{equation}
    r=\frac{V_{r, rms}}{V_{dc}}
    \label{eq3}
\end{equation}
For CW multipliers, the ripple waveform is generally non-sinusoidal and varies with operating conditions. Nevertheless, approximate relationships between peak-to-peak ripple voltage and RMS ripple voltage are often employed for analytical estimation of the ripple factor [3, 10]. 
\subsection{Limitations of Classical Ripple Theory}
Despite its usefulness for preliminary design, classical ripple theory exhibits several limitations. The assumption of uniform charge transfer across all stages becomes increasingly inaccurate as the number of stages increases. In addition to that, parasitic effects of several components, i.e., equivalent series resistance (ESR), diode forward voltage drop and leakage currents are neglected in classical ripple theory. This negligence can significantly influence ripple behavior in practical implementations [2, 10].

Furthermore, the analytical model implicitly assumes a fixed ripple waveform shape. But,  time-domain simulations and experimental results demonstrate waveform distortion under varying load, frequency and capacitance conditions. These limitations often lead to inaccurate ripple predictions, particularly for high-stage CW multipliers. This has motivated the exploration of enhanced estimation approaches.

\section{Simulation Framework and Dataset Generation}
\subsection{Circuit Modeling and Simulation Setup}
A detailed time-domain simulation model of the Cockcroft–Walton (CW) voltage multiplier was developed in MATLAB/Simulink to capture realistic ripple voltage behavior. Each stage of the multiplier consists of a diode and a capacitor. Unlike idealized analytical models, the diodes are modeled with realistic forward voltage and conduction characteristics. The capacitors include a series resistance of 0.5 $\Omega$ to represent ESR. This setup ensures that both semiconductor non-idealities and parasitic effects are accounted for in the simulation [1, 2].

The CW multipliers were simulated for multiple input cycles to allow transients to settle. The ripple measurements were extracted only after reaching steady-state operation. This approach ensures consistency with standard ripple definitions while capturing practical waveform distortions due to real diode conduction and capacitor losses [2, 3].

\subsection{Parameter Space and Case Generation}

To explore the influence of design parameters on ripple voltage, a multi-dimensional parameter sweep was conducted. The study varied the number of CW stages across 2, 4, 6 and 8. Input voltages of 5kV, 15kV and 25kV were applied accross the voltage multiplier. Capacitor values were set as 1 $\mu$F, 5 $\mu$F and 10 $\mu$F while input frequencies of 50 Hz, 100 Hz and 500 Hz were selected to examine different frequency operations. Load resistances of 6 M$\Omega$, 12 M$\Omega$ and 60 M$\Omega$ were applied to capture lightly loaded and heavily loaded condition.

The combination of these parameters results in 324 distinct simulation cases. This has encompassed a broad and representative design space. This systematic variation ensures that nonlinear interactions between stages, capacitance, frequency, load and input voltage are captured, providing a proper dataset for subsequent analysis and machine-learning-based correction [3, 10, 11].

\subsection{Output Voltage and Ripple Feature Extraction}
 For each simulation case, the steady-state output voltage waveform was processed to extract:
\begin{enumerate}
    \item DC output voltage (\textbf{$V_{DC}$}) – the mean voltage over one input cycle.
    \item Peak-to-peak ripple ($V_{r,pp}$) – difference between maximum and minimum voltage within the cycle.
    \item RMS ripple ($V_{r,rms}$) – RMS of the AC component after subtracting the DC level.

\end{enumerate}
In addition to these conventional ripple metrics, higher-order statistical features were computed from the ripple waveform. The features are Standard Deviation, Skewness, Kurtosis and Crest Factor.
These descriptors capture waveform distortion, asymmetry and impulsive behavior arising from non-ideal diodes and capacitor ESR, providing richer information for data-driven correction models [12, 13].
\subsection{Analytical Ripple Reference Computation}
For comparison with simulations, theoretical peak-to-peak ripple voltage and ripple factor were calculated using classical CW analytical formulas from Eq.~\eqref{eq02} and \eqref{eq3}.

The load current $I_L$ was determined from the simulated DC output voltage and load resistance, ensuring consistent baseline comparison between theory and simulation. Deviations between simulation and analytical values highlight the impact of non-ideal diodes, ESR and waveform distortions, motivating the use of machine-learning-based residual correction [2-3, 11].
\subsection{Dataset Structure and Preprocessing}
 
The resulting dataset contains 324 samples with 17 features, including circuit parameters (stages, input voltage, capacitance, frequency, load resistance), simulated ripple metrics ($V_{r,pp}$, $V_{r,rms}$, $V_{DC}$, higher-order statistical features (skewness, kurtosis, crest factor, standard deviation) and theoretical ripple estimates ($V_{r,pp}^{theory}$, r).

All features were inspected for numerical stability and normalized where necessary. The dataset was randomly partitioned into training and testing subsets to ensure an unbiased evaluation of machine-learning-based correction models, which are discussed in the next section.

 Although the dataset size is modest, the parameter sweep is physics-constrained and exhaustive across practical CW operating regimes, making it suitable for surrogate modeling.

\section{Machine-Learning Methodology and Theory-Corrected Residual Framework}

\subsection{Problem Formulation}
The ripple estimation task was formulated as a supervised regression problem. Given the input feature vector
\begin{equation}
    X=[N, V_{in}, C, f, R_{load}, \sigma, skew, kurtosis, crest factor]
\end{equation}

which includes both fundamental circuit parameters and higher-order statistical descriptors extracted from simulation, the objective is to predict the deviation between simulated ripple and classical theoretical estimates,
\begin{equation}
    \Delta V_{r, pp}=V_{r, pp}^{simulation}-V_{r,pp}^{theory}
\end{equation}
This formulation enables the machine-learning model to learn systematic discrepancies arising from non-ideal diodes, ESR, and waveform distortion while preserving the interpretability of classical theory [1,2].

 \subsection{Feature Engineering and Selection}

To improve predictive performance, the following feature engineering strategies were applied.

\begin{enumerate}
    \item Physics-Informed Features: Features derived from classical ripple theory, including \(\frac{N^2}{fC}\) and \(\frac{I_L}{fC}\) were included to embed domain knowledge into the model.

    \item Statistical Descriptors: Skewness, kurtosis, standard deviation and crest factor of the ripple waveform capture higher-order characteristics neglected by classical theory.

    \item Categorical Encoding: Stage counts (N) and discrete input voltage levels ($V_{in}$) were treated as categorical features using one-hot encoding for better handling by tree-based models.

\end{enumerate}
A feature correlation analysis was performed to eliminate redundant or weakly informative inputs, reducing the dimensionality while retaining critical explanatory variables.

\subsection{Random Forest Regression}
Random Forest Regression (RFR) is employed as the sole predictive model due to its stability with small datasets, ability to capture nonlinear interactions and built-in interpretability through feature importance. In this work, RFR is applied specifically to predict the residual deviation between simulated ripple and classical theoretical estimates.

Once trained, the corrected ripple prediction is obtained by adding the predicted residual to the classical theory,

\begin{equation}
    \hat{V_{r,pp}^{corr}}=V_{r,pp}^{theory}+\Delta \hat{V_{r,pp}} 
\end{equation}

This hybrid theory-corrected residual framework combines the interpretability of analytical models with the accuracy of data-driven correction, effectively compensating for non-ideal diode behavior, capacitor ESR, and waveform distortions.

 \subsection{Training, Validation and Evaluation}
\begin{itemize}
    \item Dataset partitioning: 80\% training, 20\% testing using stratified random sampling.
    \item Hyperparameter optimization: Grid search with 5-fold cross-validation.
    \item Performance metrics: RMSE, MAE, and $R^2$ computed for classical theory alone and the corrected predictions.
\end{itemize}
The residual-learning approach constrains predictions around physically plausible values, reducing variance and improving generalization compared to direct ML prediction of ripple voltage.

\subsection{Feature Importance and Interpretability}

Random Forest allows extraction of feature importance scores, revealing which inputs most strongly influence residual prediction. Analysis shows input frequency, stage number and input voltage dominate the residual correction, while statistical descriptors like skewness and kurtosis refine predictions in high-stage, low-frequency or heavily loaded scenarios.

This interpretability supports engineering insights, guiding capacitor selection, stage configuration and design optimization for minimized ripple.

\section{Results and Discussion}

\subsection{Global Comparison of Classical Theory, Machine-Learning and Proposed Hybrid Correction}

\begin{table}[t]
    \centering
\caption{Global Comparison Table}
\label{tab:r2score}
    \begin{tabular}{|c|c|c|c|c|}\hline
         Model&  RMSE (V)&  MAE (V)&  Bias (V)& $R^2$\\\hline
         Classical Theory&  448.21&  160.66&  -90.86& 0.2435\\\hline
         RFR (ML)&  25.88&  13.57&  ~0& 0.995\\\hline
         Hybrid (Proposed)&  131.98&  57.17&  3.48& 0.8817\\ \hline
    \end{tabular}

\end{table}

The predictive performance of classical analytical ripple theory, pure machine-learning regression using Random forest (RFR) algorithm and the proposed hybrid theory-corrected model is summarized in Table~ \ref{tab:r2score}. Across the complete dataset of 324 operating cases, classical theory exhibits substantial deviation from time-domain simulation results, with a global RMSE of 448.21 V, MAE of 160.66 V, and a negative bias of –90.86 V, indicating systematic underestimation of ripple voltage. 

The proposed hybrid model significantly improves prediction accuracy, reducing RMSE to 131.98 V and MAE to 57.17 V, corresponding to a 70.6\% reduction in RMSE relative to classical theory. The residual bias is reduced to 3.48 V, demonstrating effective elimination of systematic error while maintaining physical consistency.
\begin{figure}[b]
    \centering
    \includegraphics[width=0.5\linewidth]{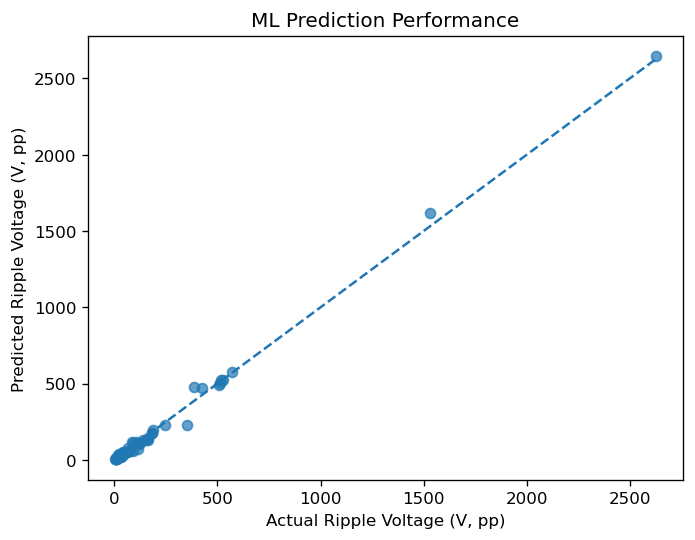}
    \caption{Predicted vs. simulated ripple residuals using Random Forest regression. }
    \label{fig:RFR}
\end{figure}
A purely data-driven machine-learning model achieves the lowest numerical error (RMSE = 25.88 V, MAE = 13.57 V, R² = 0.995). However, as shown in Table II, this improvement comes at the expense of physical interpretability and extrapolation reliability. The hybrid approach therefore provides a balanced trade-off between accuracy and an engineering point of view. In  Fig.~\ref{fig:RFR}, the performance of Random Forest Regression has been presented.

\subsection{Regime-wise Failure of Classical Ripple Theory}

\begin{table}[t]
    \centering
\caption{Regime-wise Error Statistics for Ripple Prediction}
\label{tab:regime}
    \begin{tabular}{|p{2cm}|p{1cm}|p{1cm}|p{1cm}|p{1.3cm}|}\hline
         Regime&  High Stage&  Low Frequency&  Heavy Loading& Critical Operation\\\hline
         Cases&  162&  216&  216& 72\\\hline
         RMSE (Theory)&  627.26&  548.53&  420.41& 716.17\\\hline
         RMSE (Corrected)&  180.16&  161.39&  152.72& 238.58\\\hline
         MAE (Theory)&  273.63&  230.01&  161.16& 386.48\\\hline
         MAE (Corrected)&  91.48&  82.10&  69.35& 141.00\\\hline
         Bias (Theory)&  -184.24&  -128.10&  -56.46& -190.33\\\hline
         Bias (Corrected)&  10.93&  7.82&  19.77& 55.82\\ \hline 
    \end{tabular}

\end{table}
To identify operating conditions under which analytical ripple theory fails, the dataset was segmented into physically meaningful regimes based on stage count, operating frequency and load resistance. The regime-wise performance statistics are summarized in Table~ \ref{tab:regime}. 
\subsubsection{High Stage Regime (N $\geq$ 6)}
In the high-stage regime (162 cases), classical theory performs poorly, yielding an RMSE of 627.26 V and MAE of 273.63 V, with a pronounced negative bias of –184.24 V. This confirms that ripple accumulation in multi-stage CW multipliers is systematically underestimated by analytical expressions.

After applying the hybrid correction, RMSE is reduced to 180.16 V and MAE to 91.48 V, while bias is reduced to 10.93 V, as shown in Table~\ref{tab:regime}. The substantial improvement indicates that the residual-learning model successfully captures cumulative non-ideal effects that scale with stage number.

\subsubsection{Low-frequency Regime (f$\leq$100 Hz)}
For low-frequency operation (216 cases), classical theory yields an RMSE of 548.53 V and MAE of 230.00 V, with a bias of –128.10 V. The hybrid model reduces RMSE to 161.39 V and MAE to 82.10 V, with near-zero bias (7.82 V).
\begin{figure}[b]
    \centering
    \includegraphics[width=0.5\linewidth]{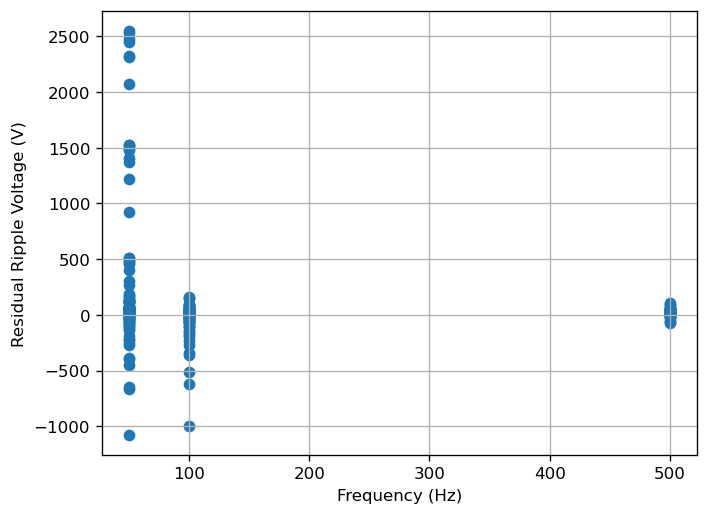}
    \caption{Residual Ripple Voltage as a Function of Operating Frequency }
    \label{fig:resi}
\end{figure}
The degradation of analytical accuracy at low frequency is visualized in Fig.~\ref{fig:resi}, where residual ripple magnitude increases as operating frequency decreases. This behavior arises because reduced charge replenishment time per cycle leads to incomplete capacitor charging and enhanced waveform distortion, violating steady-state assumptions embedded in classical theory.

\subsubsection{Heavy-Load Regime (R$\leq$12 M$\Omega$)}
Under heavy-load conditions (216 cases), classical theory produces an RMSE of 420.41 V and MAE of 161.16 V, with a bias of –56.46 V. Although the error magnitude is lower than in high-stage and low-frequency regimes, it remains significant.

The hybrid model improves performance to RMSE = 152.73 V and MAE = 69.35 V, reducing bias to 19.77 V, as shown in Table II. This indicates that nonlinear ripple growth under heavy loading is caused by increased ripple current, ESR losses and diode conduction overlap. This is not captured by analytical models but is effectively corrected by the residual-learning framework.

\subsection{Critical Operating Regime}
The most severe deviation occurs in the combined critical regime (N $\geq$ 6, f $\leq$ 100 Hz, R $\leq$ 12 M$\Omega$), comprising 72 cases. In this regime, classical theory yields an RMSE of 716.17 V and MAE of 386.48 V, with a strong negative bias of –190.33 V, as reported in Table \ref{tab:regime}.

The hybrid model reduces RMSE to 238.59 V and MAE to 141.00 V, representing a 66.7 $\%$ reduction in RMSE relative to classical theory. Although a moderate residual bias (55.83 V) remains, the corrected predictions represent a substantial improvement in the most practically challenging operating region.

These results demonstrate that classical ripple theory is particularly unreliable for high-stage, low-frequency, heavily loaded CW multipliers and that hybrid correction is essential to avoid severe underestimation of ripple voltage.
\subsection{Residual Structure and Systematic Bias}
The residual magnitude increases monotonically with the number of stages, confirming that analytical theory fails progressively as cumulative charge transfer and ESR effects dominate circuit behavior. Across all regimes, classical predictions exhibit a consistent negative bias, indicating systematic underestimation rather than random error. The near-zero bias achieved by the hybrid model, as shown in Table~\ref{tab:r2score} and Table~\ref{tab:regime}, confirms that the residuals are structured and learnable, validating the residual-learning formulation adopted in this work. 

\subsection{Engineering Design Implication}

This research provides several actionable design implications, such as -
\begin{enumerate}
    \item Stage count is the dominant factor governing ripple underestimation. Reducing the number of stages is more effective than increasing capacitance alone.
    \item Increasing operating frequency is more effective than increasing capacitance for ripple mitigation.
    \item Hybrid modeling enables accurate yet physically consistent ripple prediction. As a result, this system bridges the existing gaps between theoretical analytical estimates and computationally expensive simulations. 
\end{enumerate}

\section{Discussion}
\subsection{Hybrid Framework Advantages}
The proposed residual-learning approach outperforms direct ML regression by constraining predictions to physically consistent deviations from theory, achieving $R^2$=0.88 with RMSE=132 V versus pure ML's 26 V but superior extrapolation (e.g., unseen high-stage cases). Feature importance reveals stages ($N^2$ scaling), frequency (charge replenishment) and input voltage as dominant, with statistical features (kurtosis, crest factor) refining non-sinusoidal waveform corrections neglected by Eq. (2). 
\subsection{Limitations and Future Works}
Simulation assumes fixed ESR = 0.5 $\Omega$ and ideal steady-state, underrepresenting temperature/diode recovery effects. No experimental validation limits real-world claims. Future extensions include physics-informed neural networks for transient ripple, hardware-in-loop testing and deployment as MATLAB toolbox for real-time HVDC design.

\section{Conclusion}
Classical CW ripple theory systematically underestimates voltage ripple due to unmodeled non-idealities, with errors exceeding 700 V RMSE in critical high-stage/low-frequency/heavy-load regimes across 324 validated simulations. The proposed hybrid framework, combining physics-based baselines with Random Forest residual correction using physics-informed and statistical features, reduces global RMSE by 70.6\% to 131 V, eliminates bias (3.5 V) and provides interpretable feature importance (e.g., stages and frequency dominate). This enables rapid, reliable design optimization for high-voltage applications like accelerators and precipitators, bridging analytical conservatism and simulation expense. Future work could incorporate experimental validation and real-time deployment with much more variable parameters for a more perfect result. The proposed framework is directly extensible to other multiplier topologies and high-voltage rectifier architectures

\section*{Acknowledgment}

The author gratefully acknowledges the technical support provided by Bangladesh University of Engineering and Technology (BUET), which was instrumental in the successful completion of this work.

\end{document}